\documentclass[twocolumn,showpacs,preprintnumbers,amsmath,amssymb]{revtex4}
\usepackage{graphics}

\newcommand{\Perp}{{\!\perp}}
\newcommand{\V}[1]{{\mathbf{#1}}} 

\newcommand{\nablap}{\nabla_\Perp}

\newcommand{\sign}{{\rm sgn}}
\newcommand{\BesselK}{{\cal K}}

\newcommand{\vecv}{\V{v}}

\newcommand{\vecdn}{\delta\V{n}}

\newcommand{\ftot}{f_{\text{tot}}}

\newcommand{\denLSDplain} {\cosh f_l(x) - \cos y}

\newcommand{\Fcurv}{F_{\text{curv}}}     
\newcommand{\Fcompr}{F_{\text{compr}}}   
\newcommand{\Icurv}{I_{\text{curv}}}     
\newcommand{\Icompr}{I_{\text{compr}}}   

\newcommand{\Fi}{F_{\text{int}}} 

\begin{document}

\title{Nonlinear Effects in the TGB$_A$ Phase}
\author{Igor Bluestein}
\author{Randall D. Kamien}
\affiliation{Department of Physics and Astronomy,
   University of Pennsylvania,\\
   Philadelphia, PA 19104}
\date{\today}

\begin{abstract}
We study the nonlinear interactions in the TGB$_A$
phase by using a rotationally invariant elastic free
energy. By deforming a single grain boundary so that the
smectic layers undergo their rotation within a finite interval, we
construct a consistent three-dimensional structure.  With
this structure we study the energetics and predict the
ratio between the intragrain and intergrain defect spacing, and
compare our results with those from linear elasticity and
experiment.
\end{abstract}
\maketitle

The resolution of frustration is a central theme in condensed
matter physics. Competing terms in the free energy can favor
incompatible spatial organizations, the resolution of which often leads
to a rich phase behavior and intricate spatial patterns.  The TGB
phase of chiral smectics is an ideal example of the resolution of frustration
\cite{Lubensky1988}. While the smectic part of
the free energy favors a lamellar structure, the
chiral nematic part favors a uniform twist of the nematic
director.  These two structures are incompatible and so the layered
structure must be riddled with defects to accommodate the twist. The
TGB phase balances the competing interactions by forming a lattice
of screw dislocations arranged in twist grain boundaries.  In
earlier work \cite{TGBLinear} we investigated the geometry of the
TGB phase within a harmonic free energy in which topological
defects interact via screened exponential potentials.  However, it
was argued in \cite{KlemanBook}\ and \cite{Kamien1999}\ that
defects in the {\sl same} grain boundary interact via power-law
potentials when those nonlinearities required by rotational
invariance were included in the free energy.  In this
letter we investigate the rotationally-invariant, nonlinear energetics of the
full three-dimensional TGB$_A$ phase,
where the smectic blocks between the grain boundaries are
smectic-$A$.
We find that intergrain and intragrain interactions of
dislocations are both power-law, and we compute the aspect ratio
between the intragrain spacing $l_d$ and the intergrain spacing
$l_b$.  Our computed value is of the same order of magnitude as that computed from the linear theory and
measured by experiment.  We will comment on the discrepancy and its
source in the conclusion.

In an earlier paper \cite{TGBLinear}, we considered the screw
dislocation lattice in the TGB$_A$ phase near the transition to the nematic phase. In this regime the
dislocation density is low, so the interaction between dislocation
cores is negligible.  This implies that the lattice structure is completely
determined by the elastic energy cost of the smectic distortions
created by the screw dislocations. We demonstrated in
\cite{TGBLinear} that these distortion fields can be consistently
treated within the harmonic approximation to the elastic free
energy.  In this approximation, we express the elastic
free energy in terms of the (Eulerian) layer displacement field
$u$ and the deviation $\vecdn$ of the nematic director from the
$z$-axis:
\begin{eqnarray}\label{fsm}
  F& =& \frac{1}{2} \int d^3 x\bigg[B(\partial_z u)^2
   +D(\nablap u + \vecdn)^2\\\nonumber
    &&+K_1(\nablap \!\cdot \vecdn)^2 + K_2(\nablap \!\times \vecdn)^2
   +K_3(\partial_z \vecdn)^2\bigg],
\end{eqnarray}
where $B$ and $D$ are elastic moduli and $K_1$, $K_2$ and $K_3$
are the splay, twist and bend Frank constants, respectively. This free energy is analogous to that of a type II
superconductor and the topological defects in a smectic are likewise
analogous to Abrikosov vortices.  Employing (\ref{fsm})
the interaction energy between two parallel screw
defects is
\begin{equation}\label{fsmint}
   \frac{\Fi}{L} = \frac{D d^2}{2 \pi} \BesselK_0
   \left( \frac{r}{\lambda} \right),
\end{equation}
where $\BesselK_0$ is the modified Bessel function of order zero,
$d$ is the equilibrium layer spacing of the smectic phase, and
$\lambda \equiv\sqrt{K_2/D}$.

A twist grain boundary is composed of an array of equidistant
parallel dislocations, separated by a distance $l_d$, which force
the smectic layers to rotate. On average, the smectic layers drag the nematic
director along and allow it to twist.  As with screw defects and
vortices, the TGB phase is the analog of the Abrikosov vortex
lattice phase in superconductors \cite{Lubensky1988,deGennes1972},
although the lattice structure of defects in the TGB phase is significantly
more complicated due to the relative rotation of dislocations in different
grain boundaries. We showed in \cite{TGBLinear} that the linearized
interaction energy of two grain boundaries is
\begin{equation}
   \frac{\Fi}{A} = \frac{D d^2}{2 l_d^2} \lambda e^{-l/\lambda}.
\end{equation}
Thus the linear theory predicts an exponential interaction
between screw dislocations both within the grain boundary and
between different grain boundaries. It is simple to check that the
elastic interactions of dislocations within a grain boundary are
balanced by the interboundary interactions, which leads to a stable
ratio $R=l_d/l_b\approx 0.95$ \cite{TGBLinear}, within the experimental range
of $0.74$--$1.08$ \cite{Navailles1998,Isaert1994}.

Since both smectic and cholesteric ordering are spontaneously
broken symmetries, it is essential that our free energy be
invariant with respect to rotations. The harmonic free energy is
only invariant under infinitesimal rotations; the extra terms
needed to render the free energy (\ref{fsm}) invariant under finite rotations
lead to nonlinearities that dramatically
change the energetics of even an isolated screw dislocation
\cite{Kamien1999}.  In order to investigate these effects, we work
with a rotationally invariant theory that depends on the only
Goldstone mode in the system,  the layer displacement field $u$
\cite{Kamien1999}:
\begin{equation}\label{Fv}
   F = \frac{1}{2} \int d^3 x \left\{
        B (v_z+\vecv^2/2)^2
        +K (\nabla\!\cdot\vecv)^2
        \right\},
\end{equation}
where $\vecv\equiv\nabla u$ is the layer tilt field, which, on
average, follows the deviation of the nematic director.
The free energy (\ref{fsm}) includes director
modes and the interaction energy (\ref{fsmint})  accounts for those modes outside the
core.  In the nonlinear theory that we are now considering, these director modes are
absent.  We believe that the essential physics is unchanged but that the energetic
details will require incorporation of these modes \cite{futurework}.
Note that
we have not used the mean curvature $H=\nabla\!\cdot[(\hat z+
\vecv)/\vert\hat z +\vecv\vert]$, but rather $\nabla\!\cdot\vecv$
in the bending term.  Both expressions are rotationally invariant
and though the former has greater geometric significance
\cite{KlemanBook}, the latter is significantly easier to manage.
In the limit of low-angle grain boundaries, the two expressions
concur.  Moreover, for the energy minimizing structures we will
consider here, the mean curvature remains finite everywhere
as long as $\nabla\!\cdot\vecv$ is finite.

We construct a grain boundary as a linear superposition of
parallel equidistant screw dislocations (LSD)
\cite{Kamien1999,Kamien2001} and then use it to construct a TGB$_A$
phase by compressing each of these individual boundaries along the
pitch axis to force their rotation into a finite interval.  We may
then piece together a periodic lattice of these grain boundaries
through translation and rotation in order to construct
a full, three-dimensional TGB structure.  We will calculate
the free energy of this structure using (\ref{Fv}) to find a
minimum energy solution and consequently the geometric parameter
$R\equiv l_d/l_b$ for the system.  As noted in \cite{Kamien1999},
an important difference between the nonlinear and the harmonic theories is that in the
former the
interaction between the dislocation cores may not be neglected: the
size and shape of the core region is another variational parameter.

We start with the layer distortion field for a linear
superposition of screw dislocations (LSD) in a smectic liquid
crystal.  Through deformation this will become the displacement
field in a single smectic-$A$ block.  Choosing the average layer
normal to point along the $z$-axis, and expressing the
displacement $u$ and co\"ordinates $x$ and $y$ in units of
$l_d/2\pi$, we have \cite{Kamien2001}
\begin{eqnarray}\label{originalLSD}
   u(x,y,z) &=& -2 \sin \left(\frac{\alpha}{2}\right)\tan^{-1}
               \left[\frac{\tanh (x/2)}{\tan (y/2)}\right]\nonumber \\
               &&\qquad+(\cos \left(\frac{\alpha}{2}\right)-1)z.
\end{eqnarray}
The layer rotation angle $\alpha$ across the grain boundary is set
by the topology of the grain boundary and, therefore, does not
depend on specific details of the grain boundary model. It is
determined by the layer spacing $d$ and the dislocation spacing
$l_d$ through $\sin \left(\frac{\alpha}{2}\right) = d/2 l_d$. The
first term in (\ref{originalLSD}) describes an array of parallel
screw dislocations aligned along the $z$-direction at $x =0$, $y =
0, \pm l_d, \pm 2 l_d, \dots$, while the second term ensures that
the nonlinear compression energy completely vanishes far away from
the dislocation array.

To construct the TGB$_A$ phase out of individual twist grain
boundaries, we confine the layer rotation of the regular LSD to a
finite interval, $(-l,l)$ so that the grain boundaries are spaced
$2l$ apart in units of $l_d/2\pi$, so $2l={2\pi\over l_d}l_b={2\pi\over R}$.
To do this, we replace the co\"ordinate $x$
in (\ref{originalLSD}) with an odd deformation function, $f_l(x)$
which monotonically maps the finite interval $(-l,l)$ to
$(-\infty,\infty)$. So that we recover the single grain limit,
$f_l(x) = x$ for large $l$. The energetics of the confined
LSD will impose additional constraints on $f_l(x)$. These
constraints and the optimal choice of $f_l(x)$ will be the focus
of our calculation.  Since the energy depends on $\vecv$
we consider:
\begin{eqnarray}
\label{vx}
  v_x &=& -\sin \left(\frac{\alpha}{2}\right)\frac{\sin y}{\denLSDplain}f_l'(x),\\
\label{vy}
  v_y &=& \sin \left(\frac{\alpha}{2}\right)\frac{\sinh f_l(x)}{\denLSDplain},\\
\label{vz}
  v_z &=& \cos\left(\frac{\alpha}{2}\right)-1.
\end{eqnarray}
The compression elastic free energy per dislocation per unit
length in the $z$-direction can be found using (\ref{Fv}) and the
expressions (\ref{vx}-\ref{vz}):
\begin{eqnarray}\label{Fcompr}
 \frac{\Fcompr}{l_z} =
         2 B \sin^4 \left(\frac{\alpha}{2}\right)\left(\frac{l_d}{2 \pi}\right)^2
         \Icompr[f_l,a;R],
\end{eqnarray}
where $\Icompr$ is a dimensionless functional that depends on the
deformation function, a short distance cutoff $a(x)$ required
to excise the core from the integration domain \cite{Pleiner,Kamien1999}, and the lattice
aspect ratio $R$.
It is interesting that this energy is essentially nonlinear since
it depends on the fourth power of the displacement field $u$
\cite{KlemanBook}.

On the other hand, the curvature contribution to the free energy
per dislocation need not have a short-distance cutoff: as we will
show, by judicious choice of the deformation $f_l(x)$, the
curvature is everywhere finite.  Integrating along the grain
boundary direction ($y$) we find:
\begin{eqnarray}\label{Fsplay}
  \frac{\Fcurv}{2Kl_z}
   &=&\sin^2 \left({\frac{\alpha}{2}}\right) \pi \int_0^{\pi/R}d x
     \left\{
        \frac{1}{2} \frac{\coth f_l}{\sinh^2 f_l}[(f_l')^2-1]^2\right.\nonumber\\
        &&\left. -\frac{1}{\sinh^2 f_l}[(f_l')^2-1]f_l''+(\coth f_l - 1)(f_l'')^2
      \right\} \nonumber\\
    &\equiv & \sin^2 \left(\frac{\alpha}{2}\right) \frac{\Icurv[f_l;R]}{2K}.
\end{eqnarray}
Since a single helicoidal screw dislocation has no curvature, we
need not excise the core region in the curvature energy as long as
we maintain the screw-like character of the defect.  Near $x=0$,
$f_l(x)\approx 0$ is small and so we may expand the integrand
${\cal I}_{\rm curv}$:
\begin{equation}
   {\cal I}_{\rm curv}=\frac{\pi[(f_l')^2-1]^2}{2 f_l^3}
      -\frac{\pi[(f_l')^2-1]f_l''}{f_l^2}+\frac{\pi(f_l'')^2}{f_l}.
\end{equation}
It is evident from the above expression that the curvature energy
diverges near $x=0$ unless $f_l'(0)=1$ and $f_l''(0)=0$.
Thus we have two additional constraints on $f_l(x)$.  To finish
our analysis we must consider the behavior of the energy halfway
between grain boundaries at $x=\pm l$.  It is there also that we must take care
to match the layer normals in successive blocks so that the
nematic director is well defined everywhere.

Near $x=\pm l$, the compression energy is finite since $\vecv$ is
well behaved. However, the curvature energy is more problematic.
Since $f_l(x)\rightarrow\infty$ as $x\rightarrow l$, we can expand
the integrand in powers of $e^{-f_l}$:
\begin{equation}\label{asympSplay}
   {\cal I}_{\rm curv}=2 \pi \sum_{n = 1}^\infty e^{-2 n f_l}
   [f_l''-n (f_l')^2+n]^2.
\end{equation}
Since $f_l(x)$ diverges at $x=l$, we might consider setting the
largest term to zero, {\sl i.e.} the $n=1$ term in
(\ref{asympSplay}). Implementing the boundary conditions $f_l(0) =
0$ and $f_l(l)=\infty$, we find:
\begin{equation}\label{mapFunc0}
   f_l(x) = \sign(x)\log \left\{\frac{\sinh l}{\sinh(l-|x|)}\right\}.
\end{equation}
It is straightforward to check that as $x\rightarrow l$, the only
nonvanishing term in (\ref{asympSplay}) is the $n=2$ term and that
at $x=l$
\begin{equation}\label{asympStrain0}
  {\cal I}_{\rm curv}= \frac{2 \pi}{\sinh^4 l}.
\end{equation}
We can embellish (\ref{mapFunc0}) with a power series around
$\vert x\vert =l$ so that the first and second derivatives satisfy
the required conditions at the origin:
\begin{eqnarray}\label{fOpt}
  f_l(x) = &&\sign(x)\log \left\{\frac{\sinh l}{\sinh(l-|x|)}\right\}\\ \nonumber
         &&+\sign(x)[a_0+a_2(l-|x|)^2+a_4(l-|x|)^4],
\end{eqnarray}
where $a_0$, $a_2$ and $a_4$ may be straightforwardly calculated
\cite{footnote1}. These extra terms do not spoil the divergent
behavior at $x=l$ and so it would appear that we have found a low
curvature deformation of the original grain boundary.  However,
our goal is to assemble individual blocks into the TGB structure.
This is only possible if the smectic layers match at the midplane.
Computing the layer tilt at the midplane, we find that $v_x$ has a
small periodic component, while $v_y$ attains a constant value
corresponding to the rotated layers:
\begin{eqnarray}
  v_x(l,y) &=& -2 \sin \left(\frac{\alpha}{2}\right)\frac{\sin y}{\sinh l},\\
  v_y(l,y) &=& \sin \left(\frac{\alpha}{2}\right).
\end{eqnarray}
Since $v_x$ reflects the periodicity of a single dislocation array
and in adjacent blocks the dislocation arrays are rotated with
respect to each other, the nonvanishing of $v_x$ at the midplane
makes a perfect fit impossible. This might be a consequence of the
limited class of transformations considered in our model that
allow the smectic layers to relax only in the direction of the
pitch axis. A more general model which allows the defects in the
layers to ripple in harmony with neighboring grain boundaries
might allow for a nontrivial value of $v_x$ at $x=\pm l$.  Indeed, rippling defects
have been considered to explain commensurate twist grain boundary
structures \cite{TRL}.
\begin{figure}[htb]
\includegraphics{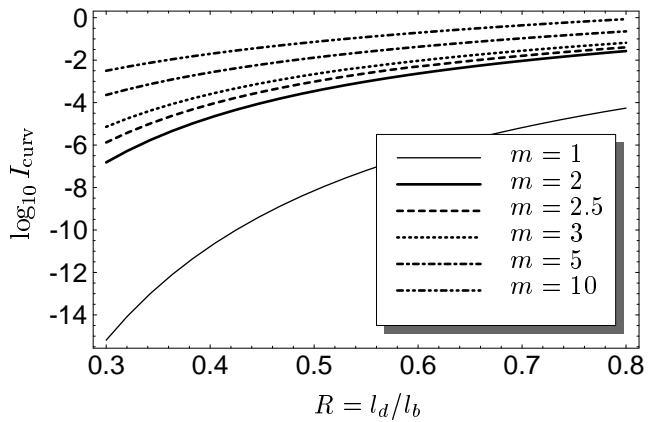} 
\caption{Dependence of $\Icurv(R;m)$ on the deformation parameter $m$.}
\label{figcurvm}
\end{figure}

The smectic layers could be matched perfectly
if $f_l$ were modified to diverge faster than the form in (\ref{mapFunc0}).
Consider a one-parameter family of deformations based on (\ref{mapFunc0}):
\begin{equation}\label{mapFuncn}
  f_l(x;m)= m \;\sign(x)\log\left\{\frac{\sinh(l/m)}{\sinh[(l-\vert x\vert)/m]}\right\}.
\end{equation}
As before, these deformations can be altered so that they satisfy
all the necessary boundary conditions.  As $\vert
x\vert\rightarrow l$, both $f''_l(x;m)$ and $[f'_l(x;m)]^2$
diverge as $(l- \vert x\vert)^{-2}$, while $\exp\{-2f_l(x;m)\}
\sim (l- \vert x\vert )^{2m}$.  Examining ${\cal I}_{\rm curv}$,
we see that as long as $m\ge 2$, the curvature remains finite in
the whole region.  In addition, $v_x\sim (l- \vert x\vert)^{m-1}$,
and so for $v_x$ to vanish at $x\pm l$, we must have $m>1$. By
numerically evaluating $\Icurv$, we find that the curvature energy
is an increasing function of $m$, as shown in Figure
\ref{figcurvm}. Thus we choose $m=2$ in our calculation of the
aspect ratio $R$.  We note that the isolated $m=1$ deformation has a still
lower energy, though it is, unfortunately, not allowed
because of geometry.  In future work we will reconsider the $m=1$
deformation by allowing for more general variations of the
dislocation lattice.
\begin{figure}[hbt]
  \includegraphics{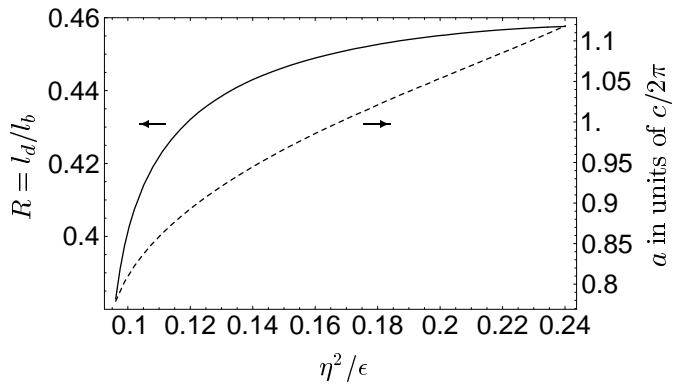} 
  \caption{Dependence of the optimal lattice aspect ratio $R$
   and core size $a$ on the control parameter $\eta^2/\epsilon$ for $m$=2.}
  \label{OptimalALGraph}
\end{figure}
The equilibrium lattice configuration is determined by minimizing the
total free energy density with respect to the dislocation spacing,
core size, and parameters characterizing the core shape. The total
energy density includes the elastic energy we have discussed, an
energetic cost from the dislocation cores and the crucial chiral
contribution which favors twist. We assume that the core energy is
proportional to the cross-sectional area $A$ of the core region,
with an energy per unit area $E$. The chiral energy gain per
dislocation $H$ is independent of the details of the dislocation
arrangement, and so for given values of $E$ and $H$ we must
minimize
\begin{eqnarray}
   \ftot &=& \frac{1}{l_d l_b}
         \bigg[
             2 B \sin^4 \left(\frac{\alpha}{2}\right)\left(l_d/2 \pi \right)^2
             \Icompr(A,R) \nonumber\\
            &&+2 K \sin^2 \left(\frac{\alpha}{2}\right) \Icurv(R) + E A - H
         \bigg]
\end{eqnarray}
with respect to $l_d$, $l_b$, and $A$. Note that the integrals
appearing in the compression and layer curvature terms do not
depend explicitly on the layer rotation angle $\alpha$. It is
convenient to separate the effects of varying the dislocation
density $1/(l_d l_b)\equiv 1/c^2$ and the lattice aspect ratio $R
\equiv l_d/l_b$. In terms of $R$ and $c$, $l_d = c \sqrt{R}$ and
$l_b = c/\sqrt{R}$. For simplicity, we assume that dislocation
core regions are square with sides $a$ in units of $c/2\pi$.  Our numerical
investigation of the $\Icompr[f_l(x),a(x);R]$ suggest that the
core, though essential, does not greatly alter the energetics \cite{futurework}.
In
computing $\Icompr$, the core size has to be re\"expressed in units
of $l_d/2 \pi$, so that it does not vary as $l_d$ is varied.
In units of $l_d/2\pi$, the core size is $a/\sqrt{R}$.  The entire minimization procedure can
be formulated in terms of the dislocation density $1/c^2$, the
aspect ratio of the dislocation lattice $R$, and the core size
$a$. Recalling that $ \sin \left(\frac{\alpha}{2}\right) =d/2 l_d$ and assuming $K/B= d^2$, we have
\begin{eqnarray}\label{fLattice}
   \ftot &= &\frac{B d^4}{32 \pi^2 }
         \bigg[
             \frac{1}{c^4 R}
             \Icompr(a/\sqrt{R},R)\nonumber\\
           &&\qquad+\frac{16 \pi^2}{c^4 R} \Icurv(R)
              +\frac{\epsilon a^2}{4\pi^2}- \frac{\eta}{c^2}
         \bigg],
\end{eqnarray}
where $\epsilon\equiv(32 \pi^2/B d^4)E$ and $\eta\equiv (32 \pi^2/B
d^4)H$.  When we minimize $\ftot$ with respect to $c$, $a$ and $R$, we find that the equations
for the optimum values of $R$ and $a$ only depend on the combination $\eta^2/\epsilon$. The optimal values of
$R$ and $a$ computed for $m = 2$ are given in Figure
\ref{OptimalALGraph}.

We find that the variation in the lattice aspect ratio mainly occurs for
large core energies and then asymptotes rapidly to $R\approx 0.46$ (see Figure \ref{OptimalALGraph}).
In comparison
with the linearized theory which predicts that $l_d\approx l_b$, the model
studied here predicts that the repulsion between grain boundaries is stronger than the repulsion
between the defects in the same grain boundary and hence $l_b$ is roughly twice as large
as $l_d$.  This may be a consequence of our locking the director to the layer normal
near the defect cores \cite{Pleiner}.  Indeed, from Figure \ref{OptimalALGraph} we see that the optimal
cores are rather large, about $1/6$ the defect spacing.  In future work we will
reintroduce the director modes in a rotationally invariant fashion.  This will certainly
lower the overall energy and should allow the cores to shrink.  As a final aside, we note
that were we to choose $m=1$ in (\ref{mapFuncn}) and simply ignore the director
mismatch, we would find $R    \approx 0.85$.  That solution will be studied in further work \cite{futurework}.

It is a pleasure to acknowledge stimulating discussions with
T.C.~Lubensky and L.~Navailles.  This work was supported by
NSF Grants DMR01-29804 and INT99-10017, and by a gift
from L.J.~Bernstein.


\begin{references}
\bibitem{footnote1}
We find that  $a_0= \frac{l}{8} (5 - 5 \coth l - l/\sinh^2 l)$,
$a_2 = -\frac{1}{4 l}(3 - 3 \coth l - l/\sinh^2 l)$, and $a_4 =
\frac{1}{8 l^3}(1 - \coth l - l/\sinh^2 l)$.  Note that as
$l\rightarrow\infty$ all of these coefficients vanish.
\end{references}
\end {document}